\documentclass[aps,prl,reprint,superscriptaddress,letterpaper,twocolumn,longbibliography]{revtex4-1}
\usepackage[english]{babel}
\usepackage{ucs}
\usepackage[utf8x]{inputenc}
\usepackage{amsmath}
\usepackage{mathtools}
\usepackage{amssymb}
\usepackage{amsthm}
\usepackage{mathrsfs}
\usepackage{graphicx}
\usepackage{bm}
\usepackage{bbm}
\usepackage{empheq}
\usepackage{cases}
\usepackage{euscript}
\usepackage[usenames, dvipsnames, x11names]{xcolor}
\usepackage[colorlinks=true,linkcolor=SpringGreen4,citecolor=blue,urlcolor=Magenta3]{hyperref}
\usepackage{enumitem}
\newcommand{\mm}[1]{\mathrm{#1}}
\newcommand{\abs}[1]{\left|#1\right|}

\newcommand{\di}[1]{\mathop{}\!\mathrm{d} #1}

\newcommand{\avg}[1]{\langle #1 \rangle}
\def \uc{\mathrm{c}}

\def \ud{\mathrm{d}}

\def \uR{\mathrm{R}}

\def \ha{\hat{a}}
\def \hb{\hat{b}}
\def \hd{\hat{d}}
\def \hbe{\hat{\beta}}

\def \hH{\hat{H}}

\DeclareFontFamily{OT1}{pzc}{}
\DeclareFontShape{OT1}{pzc}{m}{it}{<-> s * [1.10] pzcmi7t}{}
\DeclareMathAlphabet{\mathpzc}{OT1}{pzc}{m}{it}

\begin{document}

\title{Kinetics of Many-Body Reservoir Engineering}

\author{Hugo Ribeiro}
\affiliation{Max Planck Institute for the Science of Light, Staudtstraße 2, 91058 Erlangen, Germany}

\author{Florian Marquardt}
\affiliation{Max Planck Institute for the Science of Light, Staudtstraße 2, 91058 Erlangen, Germany}
\affiliation{Institute for Theoretical Physics, Department of Physics, University of Erlangen-Nürnberg, Staudtstrasse 7, 91058
Erlangen, Germany}

\begin{abstract}
Recent advances illustrate the power of reservoir engineering in applications to many-body systems, such as quantum simulators
based on superconducting circuits. We present a framework based on kinetic equations and noise spectra that can be used to
understand both the transient and long-time behavior of many particles coupled to an engineered reservoir in a number-conserving
way. For the example of a bosonic array, we show that the non-equilibrium steady state can be expressed, in a wide parameter
regime, in terms of a modified Bose-Einstein distribution with an energy-dependent temperature.
\end{abstract}

\maketitle

\noindent\textit{Introduction ---} 
Reservoir engineering~\cite{poyatos1996,plenio2002,kapit2017} is used to deliberately generate some desired dissipative dynamics,
as demonstrated in a variety of platforms:~atoms~\cite{krauter2011}, superconducting
circuits~\cite{murch2012,shankar2013,leghtas2015}, ion traps~\cite{barreiro2011,lin2013,kienzler2015}, and
optomechanics~\cite{wollman2015,pirkkalainen2015}. In the future, it could become particularly useful for controlling quantum
many-body systems, as theoretically proposed in several works (see, e.g., Refs.~
\cite{diehl2008,verstraete2009,cho2011,quijandria2013,kapit2014,reiter2016}). 

First experimental realizations of many-body reservoir engineering are starting to appear:~it has been used to stabilize a circuit
QED based Mott insulator in a 1D chain of eight transmon qubits against photon losses~\cite{ma2019} as well as to dissipatively
prepare quantum states in a three-transmon array~\cite{hacohen-gourgy2015}. In that way, reservoir engineering contributes to the
implementation of quantum simulators, especially in cases where the naturally available dissipation would not drive the system to
the right many-body ground state. 

In this context, one very important scenario concerns the case where the coupling to the reservoir conserves the total number of
particles~\cite{hacohen-gourgy2015}. Only then is it possible to cool without ending up in the vacuum state. Here, we are
interested in providing a general framework to quantitatively describe the kinetics of many particles being scattered among states
thanks to the (weak) interaction with an arbitrary non-equilibrium reservoir. The non-equilibrium nature is significant:~even the
steady-state distribution will depend on details of the interaction and of the reservoir noise spectrum, in contrast to the case
of a thermal heat bath encountered, e.g., in certain approaches on particle-conserving photon
equilibration~\cite{klaers2010,schmitt2014,hafezi2015}.

We will introduce a  perturbative approach to derive the steady-state distribution in momentum space. For the bosonic case treated
here, this turns out to be a ``deformed'' Bose-Einstein distribution with an energy-dependent effective temperature. Illustrating
the general theory in the specific case of a linear array of (harmonic) bosonic modes, we furthermore observe features such as
negative-temperature states and prominent accumulation of particles at certain momenta during the time-evolution. 

The physics we encounter is partially reminiscent of cavity optomechanics~\cite{aspelmeyer2014}, except that we are now dealing
with a many-particle system instead of a mechanical resonator. Furthermore, the coupling to that system has been engineered to
preserve the number of excitations, somewhat analogous to the unconventional quadratic coupling encountered in some optomechanical
systems. 

\begin{figure}[t!]
	\includegraphics[width=\columnwidth]{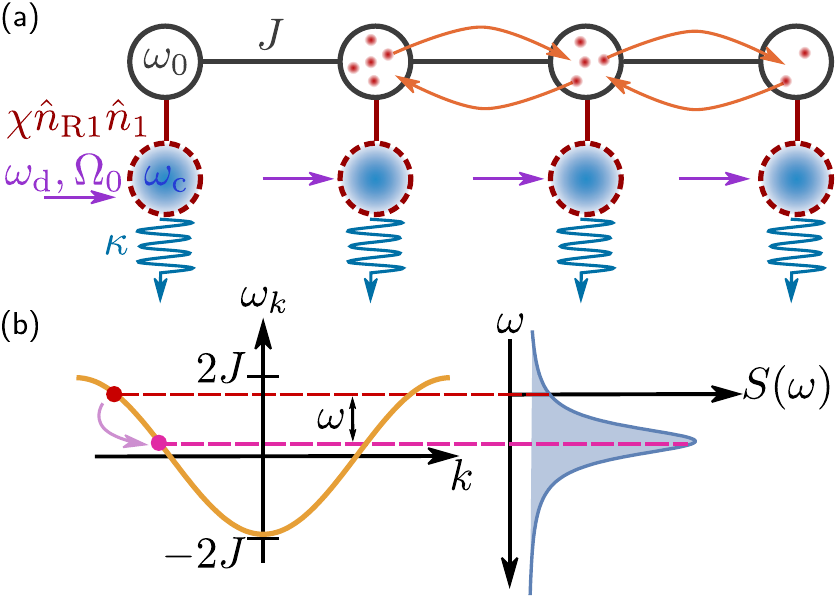}
	\caption{(Color Online) (a) Schematic representation of an array of bosonic modes, where each site experiences
		density-density coupling to a non-equilibrium reservoir (a driven, lossy cavity). (b) A particle at energy
		$\omega_1$ is scattered to a state of energy $\omega_2 < \omega_1$ by emitting a photon into the reservoir.  The
		rate at which particles scatter is proportional to the noise spectrum $S(\omega)$ of the reservoir.
	} 
        \label{fig:fig01}
\end{figure}

\noindent\textit{Model---} 
For concreteness we focus on bosons in a 1D array of modes with nearest-neighbor hopping. This physical situation can be realized,
e.g., using linear elements of the superconducting circuits toolbox (capacitively coupled cavities). Each mode is further coupled
to a lossy driven cavity via a density-density interaction [see Fig.~\ref{fig:fig01}~(a)]. The driven, lossy cavities will act as
an engineered non-equilibrium reservoir that mediates transitions between the eigenstates of the array while simultaneously
conserving the number of excitations; this ensures a well-defined chemical potential. We note that this type of coupling between
system and reservoir can be achieved, e.g., via the anharmonic transmon terms when considering a single cavity coupled
dispersively to an array of transmons. That was realized in Ref.~\cite{hacohen-gourgy2015}, using the reservoir to cool a
Bose-Hubbard chain in the atomic limit, i.e., a three-site chain with at most two excitations. An alternative implementation would
be an optomechanical array consisting of a chain of mechanical modes, each of which is coupled quadratically~\cite{sankey2010} to
a driven cavity.

The Hamiltonian describing the system is given by
\begin{equation}
	\hH (t) = \hH_\mm{array} + \hH_\mm{int} + \hH_\mm{cav} + \hH_\mm{drive} (t),
	\label{eq:Htot}
\end{equation}
where $\hH_\mm{array} = \omega_0 \sum_{j=1}^L \hb_j^\dag \hb_j + J \sum_{j=1}^{L-1} (\hb_{j+1}^\dag \hb_j + \mm{H.c.})$ with
$\omega_0$ the frequency of the modes, $J$ the hopping strength, and $\hb_j$ the annihilation operator of mode $j$. For the lossy
cavities, the Hamiltonian is $\hH_\mm{cav} = \omega_\uc \sum_j \ha^\dag_j \ha_j$, with $\omega_\uc$ the frequency, and $\ha_j$ the
annihilation operator of a photon. Most importantly, we assume a density-density coupling between the array and the reservoir, of
the form $\hH_\mm{int} = \chi \sum_j \ha^\dag_j \ha_j \hb_j^\dag \hb_j$, with $\chi$ the coupling constant. As we show later, such
a coupling allows one to cool the array while conserving the initial number of excitations present in the system.  Alternatively,
one can consider a situation where only the first site of the array is coupled to a driven reservoir.  This setting leads
qualitatively to the same cooling dynamics. However, the situation studied here presents the advantage of being homogeneous,
having a well-defined thermodynamic limit, and the cooling power scaling with the length of the array. Finally, the classical
drive on the reservoir cavity is given by $\hH_\mm{drive} = \Omega_0 \sum_j \exp(-i \omega_\ud t) \ha^\dag_j + \mm{H.c.}$, with
$\omega_\ud$ the driving frequency and $\Omega_0$ the amplitude of the driving field.

The cooling dynamics of the array is best understood by diagonalizing $\hH_\mm{array}$ and switching to a frame rotating at the
drive frequency $\omega_\ud$. To diagonalize the Hamiltonian of the array, we introduce the mode operators $\hbe_k$ defined via
the relation $\hb_j = \sum_k \varphi_k (j) \hbe_k$ where $\varphi_k (j)$ is the mode function determined via the boundary
conditions of the problem (periodic or open). Then the interaction Hamiltonian is given by 
\begin{equation}
	\begin{aligned}
		\hH_\mm{int,\uR} = \chi \sum_{j,k,k'} \varphi_j^\ast (k) \varphi_j (k') \ha_j^\dag \ha_j \hbe_k^\dag \hbe_{k'},
	\end{aligned}
	\label{eq:HIntDiagR}
\end{equation}
from which one sees that a transition from mode $k$ to $k'$ is possible without exchanging particles with the reservoir.  We note
that for a finite-size system the wave vector $k$ has a discrete set of values, which are different for open and periodic
boundary conditions. 

To understand how such an interaction can give rise to cooling dynamics, it is useful to displace the reservoir field, i.e. $\ha_j
\to a_0 + \hd_j$, where $a_0$ is the classical amplitude of the field and the operator $\hd_j$ describes the quantum noise
affecting reservoir $j$~\cite{marquardt2007,clerk2010}. Here, we neglect the fast transient dynamics and consider the reservoir to
be in the steady state. This leads to $a_0 = i \Omega_0/(i \Delta - \kappa/2)$ with $\kappa$ the energy decay rate. In the
displaced frame the Hamiltonian of the system is 
\begin{equation}
	\begin{aligned}
		\hH_\mm{eff} &=  \sum_k \omega_k \hbe_k^\dag \hbe_k - \Delta \sum_{j=1}^L\hd^\dag_j \hd_j
		+ \chi \sum_{j,k,k'} \varphi_j^\ast (k) \varphi_j (k') \\
		&\phantom{={}} 
		\times \left( \abs{a_0}^2 + a_0^\ast \hd_j + a_0 \hd_j^\dag + \hd_j^\dag
		\hd_j\right) \hbe_k^\dag \hbe_{k'},
	\end{aligned}
	\label{eq:Heff}
\end{equation}
where $\omega_k = \omega_0 + 2 J \cos(k)$ and $\Delta =\omega_\ud - \omega_\uc$ is the detuning between the reservoir cavity and
the drive. Using Eq.~\eqref{eq:Heff} one can derive Golden rule rates describing transitions from mode $k$ to $k'$, which
generalizes optomechanical cooling ideas~\cite{marquardt2007,clerk2010} to the many-body case, 
\begin{equation}
	\Gamma_{k \to k'} = \gamma \chi^2 S (\omega_{k k'}) n_k (n_{k'} +1).
	\label{eq:ratenm}
\end{equation}
We introduce the power spectrum of the non-equilibrium reservoir noise, $S (\omega)$ $=$ $\abs{a_0}^2 \int_{-\infty}^\infty \di{t}
\exp(i \omega t) \avg{\hd (t) \hd^\dag (0)}$ $=$ $\abs{a_0}^2 \kappa/[(\omega + \Delta)^2 + (\kappa/2)^2]$ [moreover, $\omega_{k k'} =
\omega_k - \omega_{k'}$ and $\gamma = \sum_{i,j} \varphi_i^\ast (k) \varphi_i^\ast (k') \varphi_j (k) \varphi_j (k')$]. 

Now we are in a position to describe the cooling dynamics of the array, by formulating a set of coupled kinetic equations that
describe the evolution of the average occupation of the eigenmodes of the array: 
\begin{equation}
	\begin{aligned}
		\dot{n} (\omega_k, t) = \sum_{k' \neq k}\left[ \Gamma_{k' \to k} (t) - \Gamma_{k \to k'} (t) \right]
	\end{aligned}
	\label{eq:KineticEq}
\end{equation}
with $n (\omega_k, t) = \avg{\hbe_k^\dag (t) \hbe_k (t)}$. In the presence of weak interactions between particles, one would have
to supplement Eq.~\eqref{eq:KineticEq} by addition of two-particle scattering terms~\cite{jaksch1997}.

As it can be seen from Eq.~\eqref{eq:ratenm}, the sign of the detuning $\Delta$ determines whether the noise $S(\omega)$ [see
Eq.~\eqref{eq:ratenm}] peaks at either negative ($\Delta>0$) or positive ($\Delta<0$) frequency. As a consequence, by choosing the
detuning $\Delta$ to be negative and assuming $\omega_k > \omega_{k'}$, we have $\Gamma_{k \to k'} > \Gamma_{k' \to k}$ in
Eq.~\eqref{eq:KineticEq}. This corresponds to particles in high-energy modes being preferentially scattered into low-energy modes
by having the reservoir absorb the excess of energy, leading to cooling. The reverse situation, with positive detuning $\Delta$,
means the noise pumps energy incoherently into the array, scattering particles to higher energies. Another property of
Eq.~\eqref{eq:ratenm} that influences the dynamics described by Eq.~\eqref{eq:KineticEq} is the Bose enhancement factor; the rate
at which a boson is scattered to a state with occupation $n$ is enhanced by a factor $n+1$. 

\begin{figure}[t!]
        \includegraphics[width=\columnwidth]{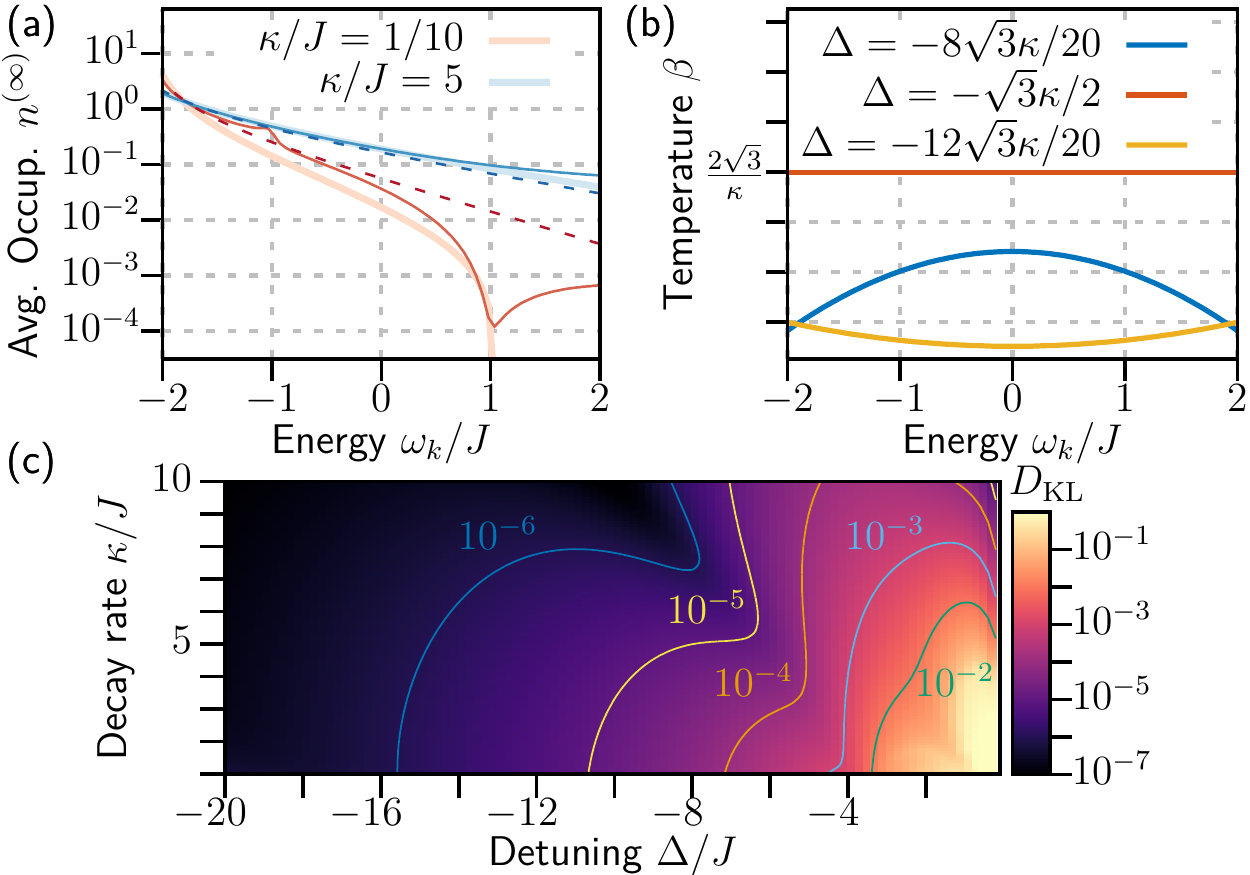}
	\caption{(Color Online) (a) Numerically observed steady-state distribution from Eq.~\eqref{eq:KineticEq} (solid thin lines),
		compared to perturbative solutions (solid thick lines) [see Eq.~\eqref{eq:ApproxSolNk}], and Bose-Einstein
		statistics (dashed lines) ($\Delta = -3J$). (b) Effective
		energy-dependent inverse temperature $\beta (\omega_k)$ of the modified Bose-Einstein distribution. (c)
		Kullback–Leibler divergence $D_\mm{LK}$ [see Eq.~\eqref{eq:KLDiv}] comparing the steady-state solution with the
		perturbative solution. In the limit $\beta_0 J \ll 1$, the perturbative solution describes accurately the steady
		state of the system. For these simulations, we considered the experimentally relevant situation of open-boundary
		conditions and the parameters $L=100$, $N=50$, $\chi = J/1000$, and $\Omega_0 = J/10$, unless specified otherwise.
	} 
	\label{fig:fig02} 
\end{figure}

\noindent \textit{Results ---}
In the long-time limit, the system approaches a steady-state distribution. In general, this does not correspond to any thermal
equilibrium distribution, as it can be seen in our numerical results [see Fig.~\ref{fig:fig02}~(a)] obtained on the basis of
Eq.~\eqref{eq:KineticEq}. An important aim in any non-equilibrium scenario is to characterize the resulting steady-state
distributions.

First insights can be obtained by recalling that even for a noise source that is not in thermal equilibrium one can always define
an effective temperature associated to a single transition frequency by using the Stokes relation $S (\omega)/S (-\omega) =
\exp[\beta_\mm{eff} \omega]$ (or, alternatively, $\tanh [\beta_\mm{eff} \omega/2] = [S (\omega) - S (-\omega)]/[S (\omega) + S
(-\omega)]$~\cite{clerk2010}). In the limit where $\Delta^2 + (\kappa/2)^2 \gg \omega^2$ we can expand $S (\pm \omega)$ in powers
of $1/[\Delta^2 + (\kappa/2)^2]$. We also expand $\tanh [\beta_\mm{eff} \omega/2]$ around $\beta_\mm{eff} \omega \to 0$. Then, we
find that the effective inverse temperature for a single transition frequency is given by
\begin{equation}
	\beta_\mm{eff} (\omega) = \beta_0\left[1 + \frac{\beta_0}{12\Delta} \left(3+\beta_0 \Delta \right)\omega^2 \right] 
	+\mathcal{O}(\omega^4)
	\label{eq:TeffNoise}
\end{equation}
where we  defined the equilibrium temperature $\beta_0 = -4\Delta/[\Delta^2 +(\kappa/2)^2]$. In the limit $\Delta  \to \infty$ or
$\kappa \to \infty$, Eq.~\eqref{eq:TeffNoise} shows that all transitions have the same temperature $\beta_0$. In this limit, the
reservoir effectively acts as a thermal reservoir and we expect the steady state to be described by Bose-Einstein statistics with
an inverse temperature $\beta_0$ and a chemical potential $\mu$: $n_\mm{BE} (\omega_k) = [\exp\left[ \beta_0 (\omega_k -\mu)
\right] - 1]^{-1}$.

We will now demonstrate that, in the vicinity of this limiting case, the steady-state distribution can be understood as a certain
deformed version of the Bose-Einstein distribution; in other words we look for a description in terms of  small corrections around
the equilibrium statistics. To do so, we turn to a perturbative solution of Eq.~\eqref{eq:KineticEq}.

The perturbation theory is best set up by considering the continuum limit of Eq.~\eqref{eq:KineticEq} for the steady state. We
have 
\begin{equation}
	\begin{aligned}
		0 &= \int_{-2 J}^{2 J} \di{\omega_{k'}} D(\omega_{k'}) \left\{ S(\omega) n(\omega_{k'})[n(\omega_k)+1] \right.\\
		&\phantom{= = \int_{-2 J}^{2 J} \di{\omega_{k'}}}
		\left. - S(-\omega) n(\omega_k) [n(\omega_{k'}) +1] \right\},
	\end{aligned}
	\label{eq:ContinuousMod}
\end{equation}
with $\omega_{k'} = \omega_k + \omega$ [see Fig.~\ref{fig:fig01}~(b)]. Here we introduced the density of states (for the 1D array,
$D (\omega) = [ {2 \pi J \sqrt{1 - \left(\frac{\omega}{2 J}\right)^2}} ]^{-1}$ when $|\omega| \leq 2J$).

The idea of the perturbative treatment is to look for a series representation of the distribution $n(\omega_k)$: $n(\omega_k) =
\sum_i n_i (\omega_k)$, with $n_i \propto (\beta_0 \omega_k)^i$ (see supplemental information for more details). We find
\begin{equation}
	\begin{aligned}
		n (\omega_k) &= n_\mm{BE} (\omega_k) \left\{ 1 - \exp[\beta_0 (\omega_k - \mu)] n_\mm{BE} (\omega_k) \times 
		\vphantom{\frac{\beta_0^2}{36 \Delta}}\right.\\
		&\phantom{={}}
		\left. \left[ \frac{\beta_0^2}{36 \Delta}(3 +
		\beta_0 \Delta)(\omega_k^2 + 18 J^2)\omega_k - \exp(-\beta_0 \mu) c \right] \right\},
	\end{aligned}
	\label{eq:ApproxSolNk}
\end{equation}
where $n_\mm{BE} (\omega_k)$ is the Bose-Einstein statistics and $c$ is a constant of integration.  We can fix both $\mu$ and $c$
in Eq.~\eqref{eq:ApproxSolNk} by enforcing particle number conservation, $\sum_k n (\omega_k) = N$; we first obtain $\mu$ by
imposing normalization for $n_\mm{BE}$ and then extract $c$ from the same condition applied to the next order.

When we compare Eq.~\eqref{eq:ApproxSolNk} to a generalized Bose-Einstein statistics with an energy-dependent inverse temperature
(setting $\beta_0 \to \beta (\omega_k)$), we conclude, to leading order:
\begin{equation}
	\beta (\omega_k) = \beta_0\left[ 1 + \frac{\beta_0}{36 \Delta}(3+\beta_0 \Delta)(\omega_k^2 + 18 J^2)\right].
	\label{eq:EffT2ndOrder}
\end{equation}
In Fig.~\ref{fig:fig02}~(a), we display the inverse temperature as a function of energy. One can see that the curvature depends
on the ratio between the detuning $\Delta$ and the decay rate $\kappa$. We note that Eq.~\eqref{eq:EffT2ndOrder} in addition to
reduce to $\beta_0$ for a small expansion parameter [see Eq.~\eqref{eq:TeffNoise}] also reduces to $\beta_0$ at the special point
$\Delta = -\sqrt{3} \kappa/2$.

To assess the accuracy of our perturbative solution, we quantify its distance from the true steady-state solution of
Eq.~\eqref{eq:ApproxSolNk}. We employ the Kullback–Leibler divergence~\cite{kullback1951} (relative entropy)
\begin{equation}
	D_\mm{KL} \left(p^{(\infty)} \parallel p\right)= \sum_k \frac{n^{(\infty)} (\omega_k)}{N} \log\left[ \frac{n^{(\infty)}
	(\omega_k)}{n (\omega_k)} \right],
	\label{eq:KLDiv}
\end{equation}
which describes the amount of information lost when approximating one distribution by another. Here, we expressed our perturbative
solution in terms of the normalized density $p(\omega_k)=n (\omega_k)/N$ [with $N$ the total number of particles and $n
(\omega_k)$ given by Eq.~\eqref{eq:ApproxSolNk}] and compare it to the steady-state distribution $p^{(\infty)}(\omega_k) =
n^{(\infty)} (\omega_k)/N$. 

Figure~\ref{fig:fig02}~(c) shows the Kullback–Leibler divergence as a function of detuning and decay rate. The numerical results
were obtained by time-evolving the kinetic equations until the system settles into the steady state. 

\begin{figure}[t!]
        \includegraphics[width=\columnwidth]{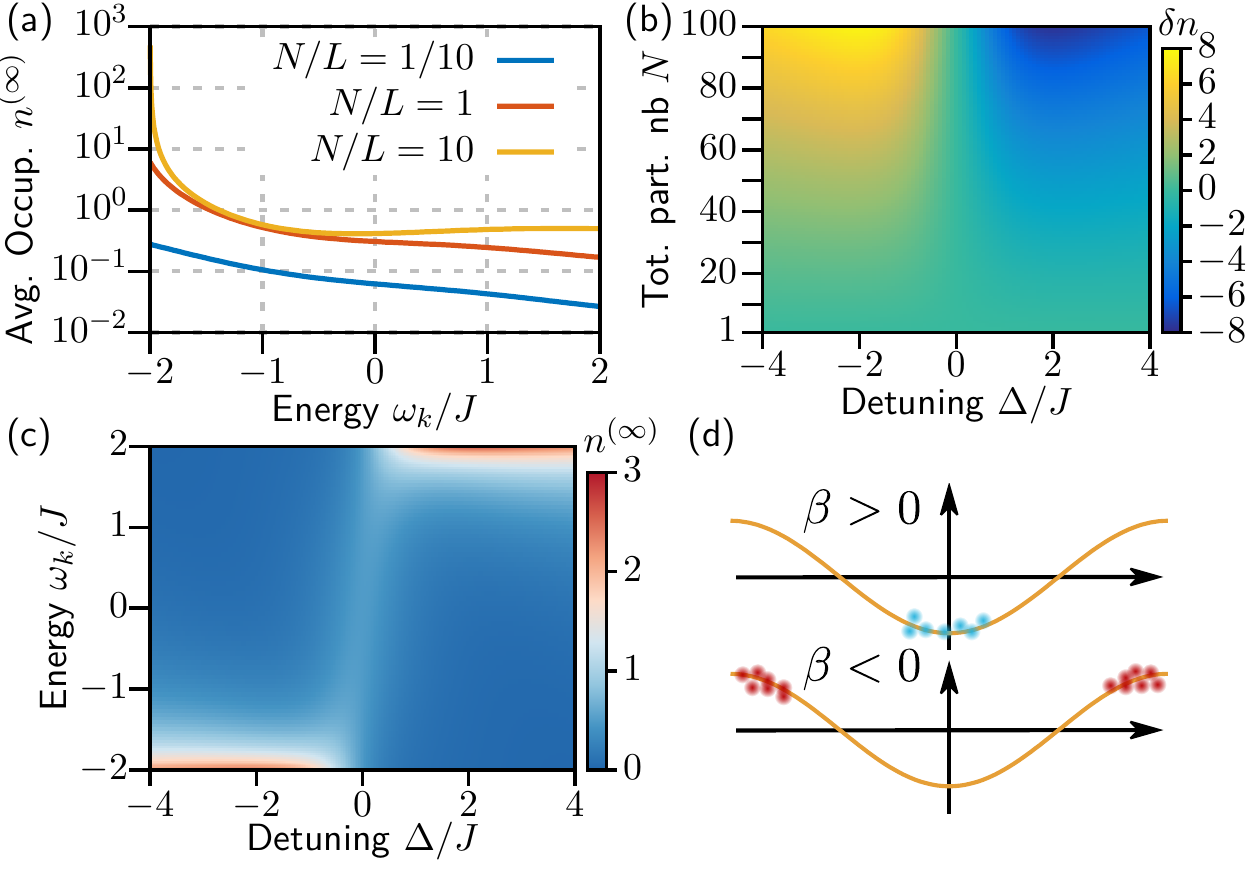}
	\caption{(Color Online) Bosonic distribution in momentum space (long-time limit). (a) Impact of the Bose enhancement
		factor ($\Delta = -J, \kappa = 3J$). As the particle density increases, the low-energy levels become more occupied
		during the cooling dynamics. (b) Efficiency of ground state cooling. We plot the difference between the average
		occupation of the ground state and the highest excited state, $\delta n = n_\mm{GS} - n_\mm{high}$ ($\kappa = 3
		J$). (c) Transition towards negative temperature. For blue detuning ($\Delta >0$), the system exhibits stable
		negative-temperatures steady states. (d) Schematic representation of states with positive and negative
		temperatures. 	
		}
        \label{fig:fig03}
\end{figure}

We observe the pertubative solution in Eq.~\eqref{eq:ApproxSolNk} approximating the non-equilibrium steady state well in a wide range
of parameters. The approximation becomes better, as expected, in the limit $\beta_0 J \ll 1$ (for $\beta_0 J\to 0$, the effective
temperature becomes energy-independent, and Eq.~\eqref{eq:ApproxSolNk} reduces back to the Bose-Einstein distribution). We stress
that Eq.~\eqref{eq:ApproxSolNk} is much closer to the real solution than the Bose-Einstein distribution itself; indeed, in a large
parameter regime we have $D_\mm{KL} (p^{(\infty)} \parallel p_\mm{BE}) / D_\mm{KL} (p^{(\infty)} \parallel p) \approx 10^2$ with
$p_\mm{BE} = n_\mm{BE}/N$ (see supplemental material).

While we focussed on the specific case of a one-dimensional chain coupled to a non-equilibrium reservoir with a particular noise
spectrum, our perturbative treatment is applicable to any spectrum and for non-interacting bosons in any dimension, in arbitrary
potentials, with suitably modified mode functions and density of states. In all of these cases, it can serve to extract the
steady-state non-equilibrium distribution brought about by reservoir-induced particle scattering. 

Equation~\eqref{eq:ApproxSolNk} represents our main analytical result. We now discuss some distinct properties of the non-equilibrium
distribution. 

In thermal equilibrium, detailed balance ensures that the Bose-Einstein distribution is independent of the density of states. This
is no longer true here, out of equilibrium. As we show in Fig.~\ref{fig:fig02}~(b), signatures of the density of states (with
its characteristic divergence at the band edge in 1D) can be observed when the noise spectrum itself [see Eq.~\eqref{eq:ratenm}]
is sufficiently narrow i.e, for $\kappa/J \ll 1$. In that case, prominent features (almost non-analytic) appear in the
distribution at energies separated from the band edge by the detuning. These features are even more pronounced during the temporal
dynamics (see below).

When deriving the kinetic equation [Eq.~\eqref{eq:ratenm}], we pointed out the presence of the bosonic enhancement factors. The
effects of such an enhancement can be observed when the particle density increases, particularly in the most strongly occupied
regions of the band [see Fig.~\ref{fig:fig03}~(a)]. For red detuning, we obtain many-particle ground state cooling. To assess its
efficiency, we plot in Fig.~\ref{fig:fig03}~(b) the difference between the occupation of the single-particle ground state and the
highest excited state in the band, $\delta n = n_\mm{high} - n_\mm{GS}$. The bosonic enhancement serves to increase the efficiency
of ground state cooling.

In stark contrast to optomechanics, where the spectrum of the mechanical resonator is unbounded, we are dealing with a physical
system that possesses a bound (many-particle) spectrum. This has an important consequence:~even for blue detuning ($\Delta>0$),
and without any recourse to non-linear cavity dynamics, the system is still stable, as can be seen in Fig.~\ref{fig:fig03}~(c). In
this ``heating'' regime, particles  accumulate at the upper band edge. Such states can be described by a negative temperature [see
Fig.~\ref{fig:fig03}~(d)]. Similar negative-temperature states were previously obtained using localized spin
systems~\cite{purcell1951,oja1997,medley2011} and ultracold atoms trapped in optical lattices~\cite{braun2013}. Creating stable
negative states with optical lattices is challenging as it requires relatively complex state preparation.  By contrast, here, one
only needs to choose a positive detuning.

\begin{figure}[t!]
        \includegraphics[width=\columnwidth]{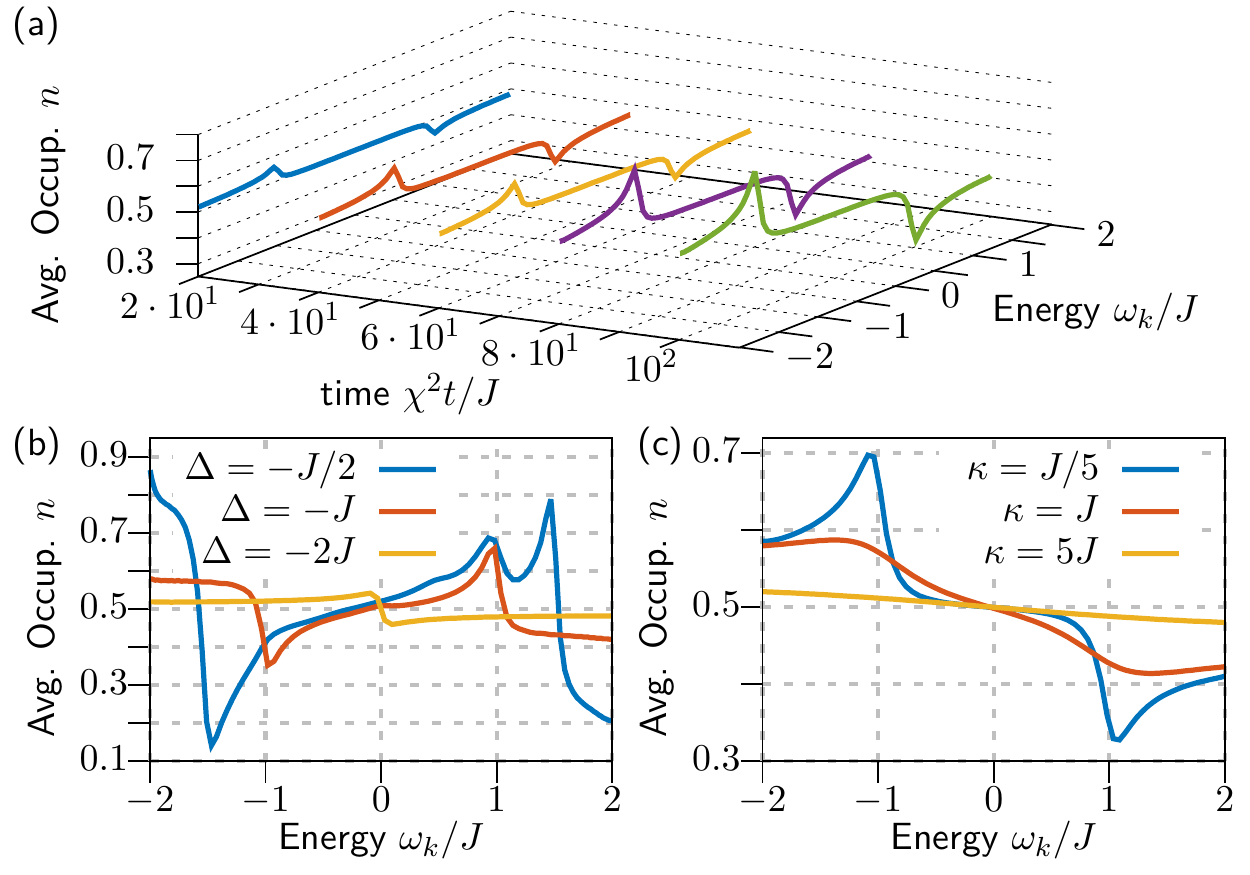}
	\caption{(Color Online) Time evolution. (a) Snapshots of the distribution showing the emergence of accumulation and
		depletion of particles at $\pm \Delta$ off the band edges ($\Delta=-3J, \kappa = J/10$). The steady-state
		distribution can be seen in Fig.~\ref{fig:fig02}~(a). (b) Detuning-dependence ($\chi^2 t/J = 10$, $\kappa=J/10$).
		(c) Impact of the decay rate $\kappa$.
	}
        \label{fig:fig04}
\end{figure}

Finally, we investigate the time-evolution. We are interested in describing the mechanism that leads to the accumulation and
depletion of particles in the distribution, as shown in Fig.~\ref{fig:fig04}. As we noted earlier, these features are present in
the steady-state distribution [see Fig.~\ref{fig:fig02}~(b)], so we expect them to be generic also during the evolution. To
consider the simplest possible situation, we assume the initial distribution to be uniform. Such a uniform distribution in
$k$-space can be realized, e.g., by incoherently loading the whole array with bosons at constant density (or quenching from a Mott
insulator). At early times, the incipient deviations from the uniform distribution can be obtained perturbatively:
\begin{equation}
	\dot{n} (\omega_k ,t) = \gamma \chi^2 n_0 (n_0 +1) \mkern-18mu \int\displaylimits_{-2 J -\omega_k}^{2 J -\omega_k} 
	\mkern-18mu \di{\omega} D(\omega +\omega_k)\left[S(\omega) - S(-\omega)\right],
	\label{eq:EarlyTimes}
\end{equation}
where we have replaced $n (\omega_k, t)$ by its initial uniform value $n_0$. The integral in Eq.~\eqref{eq:EarlyTimes} is maximal
(or minimal) when the divergence of the density of states aligns with the maximum of the spectrum $S(\omega)$ [or $S(-\omega)$].
Physically, this translates into having a large incoming rate of particles for states with $\omega_k  = 2J + \Delta$ and a large
outgoing rate for states with $\omega_k = -2J-\Delta$ (assuming $\Delta < 0$). As a consequence, particles accumulate (deplete) at
$\pm \Delta$ from the band edges [see Figs.~\ref{fig:fig04}~(b), and \ref{fig:fig04}~(c)]. The accumulation of particles at
$\omega_k = 2 J + \Delta$ can subsequently lead to another accumulation of particles at $\omega_k = 2 J + 2\Delta$ [see
Fig.~\ref{fig:fig04}~(b), $\Delta=-J/2$]. As emphasized already above for the steady state, observing these extra peaks in the
distribution is contingent on having a narrow cavity spectrum, with $\kappa \ll |\Delta|, J$ [see Fig.~\ref{fig:fig04}~(c)] 

In experimental implementations, two factors may come into play to distort the long-term dynamics from the idealized scenario - we
mention them briefly, though a detailed study would be beyond the scope of the present work. First, there is the loss of array
bosons at a constant rate (e.g. intrinsic cavity decay or mechanical dissipation). This will simply lead to an overall depletion
of the total particle number $N(t)\sim e^{-\Gamma t}$ (i.e. the steady state will slowly evolve, due to the modification of
bosonic enhancement factors). Second, if there are interactions between the bosons (e.g. attractive interactions for a chain of
transmon qubits instead of linear cavities), these will lead to additional scattering that is compatible with ground-state cooling
but will drive the distribution closer to thermal. 

\noindent \textit{Conclusion ---} We have investigated the scattering kinetics of bosonic particles subject to what is probably
the most widely used engineered reservoir, namely a driven lossy cavity. We have shown that the resulting non-equilibrium
steady-state distribution can be perturbatively analyzed in a wide parameter regime and interpreted as a deformed Bose-Einstein
distribution with an energy-dependent temperature. In this work, we have focussed on the illustrative case of a one-dimensional
array, with each site coupled to a driven cavity. However, our treatment remains valid and we expect similar observations in
general for non-equilibrium reservoirs coupled in a particle-conserving manner to arbitrary non-interacting many-body systems,
i.e.  in higher-dimensions, for arbitrary lattices and band structures, and also for fermionic systems, with straightforward
modifications.

\clearpage

\global\long\def\theequation{S\arabic{equation}}

\global\long\def\thefigure{S\arabic{figure}}

\setcounter{equation}{0}

\setcounter{figure}{0}

\setcounter{secnumdepth}{2}

\thispagestyle{empty}
\onecolumngrid

\begin{center}
{\fontsize{12}{12}\selectfont
\textbf{Supplemental  Material for
``Kinetics of Many-Body Reservoir Engineering''\\ [5mm]}}
{\normalsize Hugo Ribeiro$^1$ and Florian Marquardt$^{1,2}$\\[1mm]}
{\fontsize{9}{9}\selectfont
\textit{$^1$ Max Planck Institute for the Science of Light, Staudtstraße 2, 91058 Erlangen, Germany}\\
\textit{$^2$ Institute for Theoretical Physics, Department of Physics, University of Erlangen-Nürnberg, Staudtstrasse 7, 91058
Erlangen, Germany}}
\end{center}
\normalsize

\section{Perturbation theory}

In this section, we show in more details how we obtained Eq.~\eqref{eq:ApproxSolNk} of the main text. We start from the
equation giving the steady state in the continuum limit [see also Eq.~\eqref{eq:ContinuousMod} of the main text],
\begin{equation}
	0 = \int\displaylimits_{-2 J}^{2 J} \di{\omega_{k'}} D(\omega_{k'}) \left\{ S(\omega) n(\omega_{k'})[n(\omega_k)+1] - S(-\omega)
	n(\omega_k) [n(\omega_{k'}) +1] \right\}.
        \label{eq:ContinuousModApp}
\end{equation}
The first step consists in expanding both $S(\omega)$ and $S(-\omega)$ in powers of $\beta_0 \omega$. We have 
\begin{equation}
	\begin{aligned}
		S(\omega) &= \abs{a_0}^2 \frac{\kappa}{(\omega + \Delta)^2 + \left(\frac{\kappa}{2}\right)^2}\\ 
		&= \abs{a_0}^2 \frac{\kappa}{\Delta^2 + \left( \frac{\kappa}{2} \right)^2} \frac{1}{1 + \frac{4 \Delta}{\Delta^2 + \left(
		\frac{\kappa}{2} \right)^2} \frac{\omega}{2} \left( 1 + \frac{\omega}{2\Delta} \right) } \\
		&= \abs{a_0}^2 \frac{\kappa}{\Delta^2 + \left( \frac{\kappa}{2} \right)^2} \frac{1}{1 - \beta_0 \frac{\omega}{2} \left( 1 +
		\frac{\omega}{2\Delta} \right) } \\
		&= \abs{a_0}^2 \frac{\kappa}{\Delta^2 + \left( \frac{\kappa}{2} \right)^2} \sum_{l=0}^\infty \left[
		\frac{\beta_0}{2 \omega} \left( 1 + \frac{\omega}{2\Delta} \right)\right]^l
	\end{aligned}
	\label{eq:ExpNoiseSpectrumPos}
\end{equation}
and similarly we have 
\begin{equation}
	S(-\omega) = \abs{a_0}^2 \frac{\kappa}{\Delta^2 + \left( \frac{\kappa}{2} \right)^2} \sum_{l=0}^\infty \left[-
	\frac{\beta_0}{2 \omega} \left( 1 + \frac{\omega}{2\Delta} \right)\right]^l.
	\label{eq:ExpNoiseSpectrumNeg}
\end{equation}
We also expand $n(\omega_{k'}) = n (\omega_k + \omega)$ in a Taylor series around $\omega = 0$, we have 
\begin{equation}
	n (\omega_k + \omega) = \sum_{l=0}^\infty \frac{n^{(l)} (\omega_k)}{l!} \omega^l,
	\label{eq:ExpN}
\end{equation}
where $n^{(l)}$ denotes the $l$th derivate of $n(\omega_j)$. 

By replacing Eqs.~\eqref{eq:ExpNoiseSpectrumPos}, \eqref{eq:ExpNoiseSpectrumNeg}, and \eqref{eq:ExpN} into
Eq.~\eqref{eq:ContinuousModApp}, we can carry out the integral over the density of states. We look for solutions of the resulting
equation in the form of a series expansion
\begin{equation}
	n(\omega_k) = \sum_{l=0}^\infty n_l (\omega_k),
	\label{eq:SeriesAssump}
\end{equation}
where we assume that $n_l \propto (\beta_0 \omega_k)^l$. 

By expanding Eqs.~\eqref{eq:ExpNoiseSpectrumPos}, \eqref{eq:ExpNoiseSpectrumNeg}, and \eqref{eq:ExpN} to third order, i.e, up to
$l=3$, we can find first order differential equations for $n_l$ with $l\in\{0,1,2\}$ by grouping together terms proportional to
$(\beta_0 \omega_k)^l$. We note that we treat terms like $(\beta_0 \omega_k) \omega_k /\Delta$ as being higher order, i.e., for
this particular example this term would be grouped together with terms scaling like $(\beta_0 \omega_k)^2$.

Collecting terms proportional to $\beta_0 \omega_k$, we find that $n_0 (\omega_k)$ must obey the differential equation 
\begin{equation}
	n'_0 (\omega_k) + \beta_0 n_0 (\omega_k) [n_0 (\omega_k) + 1] = 0.
	\label{eq:n0diff}
\end{equation}
The solution of Eq.~\eqref{eq:n0} is the Bose-Einstein statistics with temperature $\beta_0$,
\begin{equation}
	n_0 (\omega_k) = \frac{1}{\exp\left[ \beta_0 (\omega_k - \mu) \right] -1}, 
	\label{eq:n0}
\end{equation}
and the constant of integration $\mu$ is the chemical potential, which is fixed by requiring that $\sum_k n_0 (\omega_k) = N$

Collecting terms proportional to $(\beta_0 \omega_k)^2$, we find that $n_1 (\omega_k)$ must obey the differential equation
\begin{equation}
	n'_1 (\omega_k) + \beta_0 n_1 (\omega_k) [n_0 (\omega_k) + 1] = 0,
	\label{eq:n1diff}
\end{equation}
whose solution is 
\begin{equation}
	n_1 (\omega_k) = c_1 \frac{\exp\left[ \beta_0 (\omega_k -2 \mu) \right]}{\left\{ \exp\left[ \beta_0 (\omega_k - \mu)
	\right] -1 \right\}^2},
	\label{eq:n1}
\end{equation}
with $c_1$ a constant of integration. To determine the constant $c_1$ we require that $ \sum_k [n_0 (\omega_k) + n_1 (\omega_k)] =
N$. Note that we have previously fixed $\mu$ by requiring that $\sum_k n_0 (\omega_k) = N$, which leads to $c_1 = 0$. We note that
this procedure to find the constants of integration ensures that our solution always describes a distribution with $N$ particles
at every order. 

Finally, collecting terms proportional to $(\beta_0 \omega_k)^3$, we find that $n_2 (\omega_k)$ must obey the differential
equation
\begin{equation}
	n'_2 (\omega_k) + \beta_0 n_2 (\omega_k) [ 2 n_0 (\omega_k) +1] + \frac{\beta_0^2 (3+\beta_0 \Delta)(6 J^2 +
	\omega_k^2)}{12 \Delta} n_0 (\omega_k) [n_0 (\omega_k) + 1] = 0.
	\label{eq:n2diff}
\end{equation}
The solution of Eq.~\eqref{eq:n2diff} is 
\begin{equation}
	n_2 (\omega_k) = - \frac{\exp\left[ \beta_0 (\omega_k - \mu \right]}{\left\{ \exp\left[ \beta_0 (\omega_k -\mu) \right]-1
		\right\}^2} \left[ \frac{\beta_0^2}{36 \Delta} (3 + \beta_0 \Delta) (18 J^2 + \omega_k^2)\omega_k - \exp(\beta_0
		\mu) c_2 \right],
	\label{eq:n2}
\end{equation}
with $c_2$ the constant of integration which is once more determined by the condition $\sum_k [n_0 (\omega_k) + n_2 (\omega_k)] =
N$.

Combining Eqs.~\eqref{eq:n0}, \eqref{eq:n1}, and \eqref{eq:n2} together with the result $c_1 =0$ leads to
Eq.~\eqref{eq:ApproxSolNk} of the main text. We note that the latter equation can predict negative occupation numbers, but this
only occurs outside of the perturbative regime where the theory is not valid anymore.

\section{Approximating the steady-state solution with the Bose-Einstein statistics}

\begin{figure}[t!]
	\includegraphics[width=0.75\columnwidth]{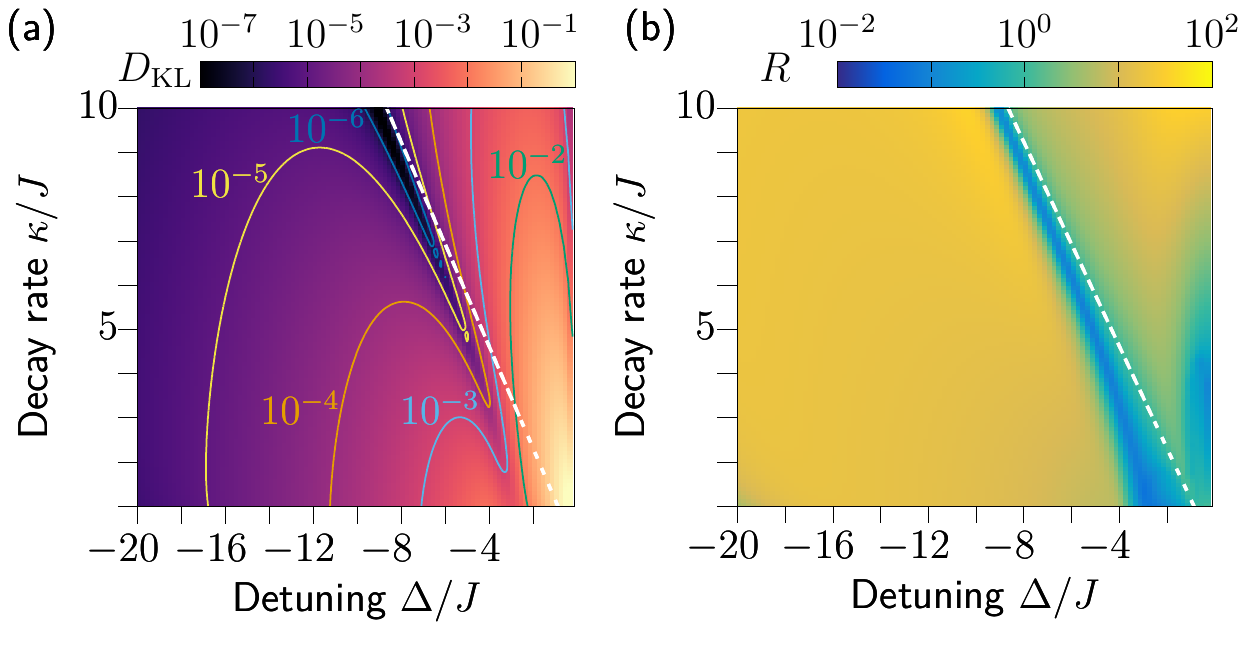}
	\caption{(Color Online) (a) Kullback–Leibler divergence comparing the steady-state solution with the Bose-Einstein
	statistics given by Eq.~\eqref{eq:n0}. (b) Ratio between $D_\mm{KL} (p^{(\infty)} \parallel p_\mm{BE})$ and $D_\mm{KL}
	(p^{(\infty)} \parallel p)$. The white dashed line corresponds to $\Delta=-\sqrt{3}\kappa/2$ where the temperature $\beta
	(\omega_k)$ is equal to $\beta_0$.}
        \label{fig:sup01}
\end{figure}

In this section we show that the third order perturbative solution is a better approximation of the steady-state solution than the
leading order Bose-Einstein statistics. In Fig.~\ref{fig:sup01}~(a) we plot the Kullback–Leibler divergence between $p^{(\infty)}$ and
$p_\mm{BE}$. We indicate by a white dashed line when $\beta (\omega_k) = \beta_0$ [see Eq.~\eqref{eq:EffT2ndOrder}], i.e., $\Delta
= -\sqrt{3}\kappa/2$. When this last condition is met, we have $n_2 (\omega_k) = 0$ if $c_2 = 0$.

In Fig.~\ref{fig:sup01}~(b), we plot the ratio
\begin{equation}
	R = \frac{D_\mm{KL} (p^{(\infty)} \parallel p_\mm{BE})}{D_\mm{KL} (p^{(\infty)} \parallel p)},
	\label{eq:rationDKL}
\end{equation}
which shows that the higher order perturbation approximates the steady-state solution more accurately than the leading order given
by the Bose-Einstein statistics. We note, however, that in the close vicinity of $\Delta=-\sqrt{3}\kappa/2$, the leading order is
more suitable to approximate the steady state.

\end{document}